\documentclass[12pt]{article}
\usepackage{amssymb}
\usepackage{graphicx}

\newcommand{\beq}{\begin{equation}}
\newcommand{\eeq}{\end{equation}}
\newcommand{\bea}{\begin{eqnarray}}
\newcommand{\eea}{\end{eqnarray}}
\newcommand{\ba}{\begin{array}}
\newcommand{\ea}{\end{array}}
\newcommand{\bi}{\begin{itemize}}
\newcommand{\ei}{\end{itemize}}
\newcommand{\bn}{\begin{enumerate}}
\newcommand{\en}{\end{enumerate}}
\newcommand{\bc}{\begin{center}}
\newcommand{\ec}{\end{center}}
\renewcommand{\l}{\left}
\renewcommand{\r}{\right}

\newcommand{\ol}{\overline}
\newcommand{\De}{\Delta}
\newcommand{\de}{\delta}
\newcommand{\be}{\beta}

\newcommand{\la}{\lambda}
\newcommand{\om}{\omega}
\renewcommand{\th}{\theta}

\newcommand{\nl}{\nonumber\\}

\newcommand{\ME}[1]{\langle{#1}\rangle}

\begin{document}
\tolerance=100000


\vspace*{\fill}

\begin{center}
{\Large \bf
Neutrino mass matrix in triplet Higgs models with $A_4$ symmetry
}\\[3.cm]

{\large\bf
Myoung Chu Oh$^a$\footnote{mcoh73@gmail.com}, ~
Seungwon~Baek$^b$\footnote{sbaek@korea.ac.kr}
}
\\[7mm]

{\it
$^a$
Department of Physics,  University of Seoul,
Seoul 130-743, Republic of Korea\\
$^b$
The Institute of Basic Science and Department of Physics,
Korea University, Seoul 136-701, Republic of Korea
}\\[10mm]
\end{center}

\vspace*{\fill}

\begin{abstract}
{\small\noindent

We consider triplet Higgs model with $A_4$ symmetry to generate the
neutrino mass matrix. The tribimaximal form of the neutrino mixing
matrix can be naturally obtained.
Imposing the neutrino oscillation data, we show that
1) both normal and inverted mass hierarchy are allowed,
2) there is a lower bound on the lightest neutrino mass
and the effective mass for neutrinoless double beta
decay,
3) the non-vanishing $\th_{13}$ can be accommodated by considering
small perturbation,
4) $\th_{\rm atm}$ should be very close to $\pi/4$ even after perturbation.
}
\end{abstract}

\vspace*{\fill}

\begin{flushleft}
\today
\end{flushleft}

\newpage

\section{Introduction}
During last decade we have seen the firm evidences that the neutrinos have
masses and mixings which may indicate the new physics (NP) beyond the
Standard Model (SM).
The solar, atmospheric and reactor neutrino experiments have measured
the mass squared differences and mixing angles~\cite{PDG}
as shown in Table~\ref{tab:osc_data}.
\begin{table}[bht]
\begin{center}
\begin{tabular}{|c|c|c|}
\hline
Parameter &  Best fit & 3-$\sigma$ c.l. range \\
\hline
\hline
  $\Delta m^2_{\odot} (10^{-5} {\rm eV}^2)$  &  $(7.65^{+0.23}_{-0.20})$ &  7.05 -- 8.34 \\
  $\Delta m^2_{\rm atm} (10^{-3} {\rm eV}^2)$ & $(2.40^{+0.12}_{-0.11})$ & 2.07 -- 2.75 \\
  $ \sin^2(\theta_{\odot}) $ & $ 0.304^{+0.022}_{-0.016}$ & 0.25--0.37 \\
  $\sin^2(\theta_{\rm atm})$ & $ 0.50^{+0.07}_{-0.06}$ & 0.36 -- 0.67 \\
  $\sin^2(\theta_{13})$ & $0.01^{+0.016}_{-0.011}$ & $\le$ 0.056 \\
  \hline
\end{tabular}
\end{center}
\caption{The neutrino oscillation data.}
\label{tab:osc_data}
\end{table}
From the data in Table~\ref{tab:osc_data} only, 
we do not know the mass hierarchy,
the absolute neutrino masses, or whether neutrinos are
Majorana or Dirac particles.
The future experiments like the neutrinoless double beta decay
or Tritium decay may give further information on the nature of neutrinos.
Therefore, it is important to get a model which {\it predicts}
the mass matrix of neutrinos.

The current experimental data for the neutrino oscillations suggest that
the mixing matrix, the Pontecorvo-Maki-Nakagawa-Sakata (MNS) matrix,
is tribimaximal at the zeroth order~\cite{Harrison}:
\begin{eqnarray}
(|U_{l \nu}|^2) = \left(
\ba{ccc}
 2/3  &  1/3 &  0 \\
 1/6  &  1/3 &  1/2 \\
 1/6  &  1/3 &  1/2
\ea
\right).
\label{eq:tribimaximal}
\end{eqnarray}
One of natural flavor symmetries which give the tribimaximal MNS
matrix is the $A_4$ symmetry~\cite{A4,Hirsch}.
In addition, it was shown that the effective mass parameter
relevant to the neutrinoless double beta decay can be predicted in a
model with $A_4$ flavor symmetry~\cite{Hirsch}.
The tribimaximal mixing can also be naturally realized in the
triplet Higgs model.

The triplet Higgs model, a TeV scale NP model,
can generate Majorana neutrino masses.
With $A_4$ symmetry the Higgs triplet model naturally
gives tribimaximal mixings to neutrinos.
The predicted doubly charged Higgs ($H^{\pm\pm}$), if light,
can be found in the 
CERN  Large Hadron Collider (LHC) experiment~\cite{Akeroyd:2007}.

In this Letter we study a triplet Higgs model with $A_4$ flavor symmetry
which predicts not only the neutrinoless double beta decay but also
all the neutrino masses.
The Letter is organized as follows:
In Section~2 we introduce our model.
The numerical analysis for the neutrino masses and mixings and the predictions
for the neutrinoless double beta decay are done in Section~3.
We conclude in Section~4.

\section{The Higgs triplet model with $A_4$ symmetry}
\label{sec:2}
The conventional method to generate the neutrino masses is to introduce the
heavy Majorana right-handed neutrinos (type-I seesaw).
However, it is difficult to test this case experimentally.
An alternative way to give Majorana masses to neutrinos is to introduce
$SU(2)_L$ Higgs triplet $\Delta$ with $U(1)_Y$ charge 1 in the SM (type-II seesaw)~\cite{HTM}.
Via Yukawa interaction the Higgs triplet model can provide  Majorana masses
to neutrinos if the neutral component $\Delta^0$ gets very small vev~\cite{Oh}.
The Yukawa interaction is given by
\bea
 {\cal L}_Y &=& {1 \over 2} h_{ij} L_{i}^T C i \tau_2 \Delta L_{j} +H.c,
\label{eq:general_Y}
\eea
where $h_{ij}$ is a complex symmetric coupling matrix, $L_{i}=
(\nu_i, l_i)^T_L$ is a
left-handed lepton doublet, $C$ is the Dirac charge conjugation operator, and
$\tau_2$ is a Pauli matrix. The Higgs triplet can be decomposed as follows:
\bea
 \Delta = \left(
 \ba{cc}
 \Delta^+/\sqrt{2} & \Delta^{++} \\
 \Delta^0 & -\Delta^+/\sqrt{2}
 \ea
 \right).
\eea
Then the neutrino mass matrix is written in terms of the vev as
\bea
  (M_\nu)_{ij} = h_{ij} \langle \Delta^0 \rangle. 
\eea

There are many attempts to obtain the form of neutrino mass matrix
suggested by Table~\ref{tab:osc_data} 
in the framework of $A_4$ symmetry in the literature~\cite{A4,Hirsch}.
The group $A_4$ has three singlet representations, ${\bf 1}$, ${\bf 1'}$, ${\bf 1''}$,
and a triplet representation, ${\bf 3}$. Their tensor products are decomposed as
\bea
 {\bf 3} \otimes  {\bf 3} = {\bf 3}_s \oplus {\bf 3}_a \oplus {\bf 1}\oplus {\bf 1'}\oplus {\bf 1''},
\quad {\bf 1'} \otimes {\bf 1''} = {\bf 1}.
\eea
For two vectors $(x_1, x_2, x_3)$ and $(y_1, y_2, y_3)$ transforming as
${\bf 3}$, the first rule in the above equation
states that
\bea
({\bf 3} \otimes {\bf 3})_{{\bf 3}_s} &=& (x_2 y_3 + x_3 y_2, x_3 y_1 + x_1 y_3, x_1 y_2 + x_2 y_1), \nl
({\bf 3} \otimes {\bf 3})_{{\bf 3}_a} &=& (x_2 y_3 - x_3 y_2, x_3 y_1 - x_1 y_3, x_1 y_2 - x_2 y_1), \nl
({\bf 3} \otimes {\bf 3})_{{\bf 1}} &=& x_1 y_1 + x_2 y_2 + x_3 y_3, \nl
({\bf 3} \otimes {\bf 3})_{{\bf 1'}} &=& x_1 y_1 + \om x_2 y_2 + \om^2 x_3 y_3, \nl
({\bf 3} \otimes {\bf 3})_{{\bf 1''}} &=& x_1 y_1 + \om^2 x_2 y_2 + \om x_3 y_3.
\eea
 We assign the three-dimensional representation
${\bf 3}$ of $A_4$ to the doublet Higgs $\phi$ and the triplet Higgs $\Delta$
as in Table~\ref{tab:qn}.
To implement $A_4$ flavor symmetric Lagrangian and get a neutrino mass
matrix consistent with the experiments, we need to introduce additional
Higgs multiplets.
Here we introduce three more triplet Higgs $\chi_{i}$ $(i=1,2,3)$
like~\cite{Hirsch}, which are assigned ${\bf 1},{\bf 1'},{\bf 1''}$ of $A_4$.
We assign ${\bf 1},{\bf 1'},{\bf 1''}$ to the right-handed leptons and
{\bf 3} to the left-handed leptons.

\begin{table}
\begin{center}
\begin{tabular}{lccccccc}
\hline
\hline
Fields & $e^c$ & $L$  & $\phi$ & $\chi_1$ & $\chi_2$ & $\chi_3$ & $\Delta$ \\
\hline
$A_4$ & $\bf{1},\bf{1'},\bf{1''}$  & $\bf{3}$ & $\bf{3}$ &$\bf{1}$ &$\bf{1'}$ &$\bf{1''}$ &$\bf{3}$ \\
$SU(2)_L$ & $\bf{1}$ & $\bf{2}$ & $\bf{2}$ &$\bf{3}$ &$\bf{3}$
&$\bf{3}$ &$\bf{3}$ \\
$U(1)_Y$ & $1$ & $-1/2$ & $1/2$ &$1$ &$1$
&$1$ &$1$ \\
\hline
\end{tabular}
\end{center}
\caption{The assignments of $A_4$ and the $SU(2)_L \times U(1)_Y$ representations in our model.}
\label{tab:qn}
\end{table}

Then not all the couplings in (\ref{eq:general_Y}) but only
the following $A_4$-symmetric Yukawa interactions
are allowed:
\bea
{\cal L} &=&
-\la_e \ol{e}_R (\phi^\dag L)_1
-\la'_e \ol{e''}_R (\phi^\dag L)_{1'}
-\la''_e \ol{e'}_R (\phi^\dag L)_{1''}
\nl
&& + {1 \over 2} \la_\Delta L^T C i \tau_2 \Delta L
+{1 \over 2} \la_1 (L^T C i \tau_2 \chi_1 L)_1
+{1 \over 2} \la_2 (L^T C i \tau_2 \chi_2 L)_{1''}
+{1 \over 2} \la_3 (L^T C i \tau_2 \chi_3 L)_{1'},
\label{eq:A4_Y}
\eea
where the subscripts $1,1',1''$ represent the
transformation rules of the $(\phi^\dag L)$ pair in the first line
and of the $(L^TL)$ pair in the second line.
The resulting lepton and neutrino mass matrices have the following form
\bea
  M_l &=& \left(
 \ba{ccc}
  \la_e v_1 &   \la_e v_2 &   \la_e v_3 \\
  \la'_e v_1 &   \om \la'_e v_2 &  \om^2 \la'_e v_3 \\
  \la''_e v_1 &   \om^2 \la''_e v_2 &  \om \la''_e v_3
 \ea
 \right), \\
  M_\nu &=& \left(
 \ba{ccc}
  a+b+c &  f   &   e \\
   f    &  a + \om^2 b + \om c & d \\
   e    &  d   & a + \om b + \om^2 c
 \ea
 \right),
\label{eq:nu_mass}
\eea
where $v_i = \ME{\phi_i} \; (i=1,2,3)$ and
\bea
&& a = \la_1 \ME{\chi_1^0}, \quad b = \la_2 \ME{\chi_2^0}, \quad c = \la_3 \ME{\chi_3^0}, \nl
&& d = \la_\Delta \ME{\Delta_1^0}, \quad
   e = \la_\Delta \ME{\Delta_2^0}, \quad
   f = \la_\Delta \ME{\Delta_3^0}.
\eea
We assume $v_1 = v_2 = v_3$, and $d = e = f$, which can be guaranteed by a residual
symmetry like $Z_3$. For simplicity we impose additional assumption: $b=c$ which
naturally gives the maximal atmospheric neutrino mixing,
although it is not required by any symmetry. It is straightforward to extend our analysis
to the case $b \not = c$. We will consider the effect of small perturbation of $c-b$ in Section~\ref{sec:bnec}.

Then the lepton mass matrix, $M_l$, can be diagonalized by rotating the left-handed lepton
by the unitary matrix
\bea
 U(\om) = {1 \over \sqrt{3}} \left(
 \ba{ccc}
  1 &  1   &   1 \\
  1  &  \om  & \om^2 \\
  1  &  \om^2   & \om
 \ea
 \right).
\eea
The neutrino mass matrix is diagonalized by the transformation
\bea
  U_\nu^T \, M_\nu \, U_\nu = M_\nu^{\rm diag},
\label{eq:eigen}
\eea
where $U_\nu$ is a unitary matrix which can be decomposed into three
successive rotation matrices: $U_\nu = V_{23} V_{13} V_{12}$~\cite{PDG}.
The form of the neutrino mass matrix when $b=c$ is given by
\bea
  M_\nu &=& \left(
 \ba{ccc}
  a+2 b &  d   &   d \\
   d    &  a -b & d \\
   d    &  d   & a -b
 \ea
 \right).
\label{eq:nu_abd}
\eea
The resulting MNS matrix is the product of $U(\om)$ and $U_\nu$:
\bea
  U_{\rm MNS} = U(\om) \, U_\nu.
\label{eq:UUU}
\eea
Alternatively we can work in the basis where the mass matrix of the charged lepton
is diagonal. In this basis the neutrino mass matrix is in the form:
\bea
  M_\nu^{\rm eff} &=& U(\om)^* M_\nu U(\om)^\dagger =\left(
 \ba{ccc}
  a+2 d &  b   &   b \\
   b    &  b   & a -d \\
   b    &  a-d & b
 \ea
 \right),
\label{eq:nu_eff} 
\eea
such that
\bea
U_{\rm MNS}^T M_\nu^{\rm eff} U_{\rm MNS} = M_\nu^{\rm diag}.
\label{eq:Mnu_eff_diag}
\eea
Since the neutrino mass matrix is symmetric under the exchange of
$\nu_\mu \leftrightarrow \nu_\tau$,
we get a bimaximal $2-3$ mixing and a vanishing $1-3$ mixing
\bea
 \theta_{23} = \pi/4, \quad \theta_{13} = 0.
\eea
The (\ref{eq:nu_eff}) does not automatically generate the desired mixing angle $\th_{12}$
for the tribimaximal mixing matrix. The form of (\ref{eq:tribimaximal}), however,
can be achieved in a wide region of parameter space.
This can be seen from the fact that the $U_{\rm MNS}$ obtained in 
(\ref{eq:Mnu_eff_diag}) can be decomposed in general as
\bea
 U_{\rm MNS}={\rm diag}(e^{i \be_1}, e^{i \be_2}, e^{i \be_3}) \;
 U_{\rm MNS}^{\rm st} \equiv P(\be) U_{\rm MNS}^{\rm st},
 \label{eq:UMNS_decomp}
\eea
where $U_{\rm MNS}^{\rm st}$ is the physically measurable matrix 
which includes the Majorana phases in the standard form~\cite{PDG},
\bea
 U_{\rm MNS}^{\rm st} = 
 \left(
 \matrix{\phantom{-}c_{12}c_{13} & s_{12}c_{13} 
       & s_{13}e^{-i\delta} \cr
   -s_{12}c_{23}-c_{12}s_{23} s_{13}e^{i\delta} 
   & \phantom{-}c_{12}c_{23}-s_{12}s_{23} s_{13}e^{i\delta} 
   & s_{23}c_{13} \cr
   \phantom{-}s_{12}s_{23}-c_{12}c_{23}s_{13}e^{i\delta} 
      &   -c_{12}s_{23}-s_{12}c_{23}s_{13}e^{i\delta} 
       &c_{23}c_{13}}  
 \right)\times {\rm diag} (e^{i\alpha_1},~
      e^{i\,\alpha_2},~1).
\eea
The phase matrix, $P(\be)$, in (\ref{eq:UMNS_decomp}) can
be absorbed into the right-handed charged lepton sector,
and therefore, is unphysical.
However, since the $M_\nu^{\rm eff}$ is complex matrix in general,
the phase matrix $P(\be)$ is generally allowed, and it makes
the allowed parameter space much more wider than the real matrix case.
The condition for the tribimaximal mixing including the unphysical
$\be$'s can be written as
\bea
 d = -\frac{(e^{i \be_1} - e^{i \be_2})[a (e^{i \be_1} + e^{i \be_2})+ b e^{i \be_2})]}
 {2 e^{2 i \be_1} + e^{2 i \be_2}},
\eea
where the phases $\be_1, \be_2$ can take arbitrary values.
The $d=0$ is a special case to give the tribimaximal mixing matrix.

We note that if we assume $\ME{\Delta_2^0}=\ME{\Delta_3^0}=0$,
the tribimaximal form can be obtained without further conditions among
$a, b, d$. This case was considered in~\cite{Ma_rapid04}.

Our model is similar to the one considered in \cite{Hirsch}. But
their neutrino mass matrix is different from ours and can be obtained
by exchanging $b \leftrightarrow d$.
Phenomenologically the two are significantly different in that,
for example, $d=0$ gives the tribimaximal mixing matrix in our case,
while the model in \cite{Hirsch} cannot.

\section{The numerical analysis and model predictions}
\label{sec:num}

\subsection{The case $b=c$}
As mentioned in the previous section, we get $\theta_{23} = \pi/4, \theta_{13}=0$
in this case. More explicitly, with $U_{\rm MNS}= U_{23} U_{13} U_{12}$, we get
\bea
 U_{23} = \left(
 \matrix{
    1 & 0 & 0 \cr
	0 & {1 \over \sqrt{2}} & -{1 \over \sqrt{2}} \cr
	0 & {1 \over \sqrt{2}} & \phantom{-}{1 \over \sqrt{2}} 
 }
 \right), \quad
  U_{13} = \left(
 \matrix{
    1 & 0 & 0 \cr
	0 & 1 & 0 \cr
	0 & 0 & 1 
 }
 \right).
\label{eq:U23}
\eea
After rotating in the $23$ and $13$ plane, we can block-diagonalize 
the $M_\nu^{\rm eff}$ as
\bea
 U_{\rm MNS}^T M_\nu^{\rm eff} U_{\rm MNS} = U_{12}^T \;
  \left(
  \matrix{
  a+2 d &  \sqrt{2} b   &   0 \cr
  \sqrt{2} b    &  a+b-d   & 0 \cr
   0    &  0 & -a+b+d}
  \right) \;
 U_{12} = M_\nu^{\rm diag}.
 \label{eq:Mnu_diag2}
\eea
We can read $m_3^2 = |-a+b+d|^2$ and obtain the remaining mass-squared 
eigenvalues, $m_1^2, m_2^2$, and the remaining mixing angle, $\th_{12}$,
by squaring (\ref{eq:Mnu_diag2}), {\it i.e.}
$U_{\rm MNS}^\dag {M_\nu^{\rm eff}}^\dag M_\nu^{\rm eff} U_{\rm MNS} 
= \l( M_\nu^{\rm diag} \r)^2$.
There can be spontaneous CP violation
and the parameters $a, b, d$ are in general complex numbers:
$a = |a| e^{i \varphi_a}, b = |b| e^{i \varphi_b}$, and
$d = |d| e^{i \varphi_d}$.
However, one of the vev's can be made real by $SU(2)$
rotation. 
So we can set $\varphi_d \equiv 0$ without loss of generality. 
Then it is straightforward to
get the mixing angle $\theta_{12}$ of $U_{12}$ and
the mass-squared eigenvalues:
\bea
 m_{1,2}^2 &=&  {|d|^2 \over 2} | \Lambda \mp \Delta |, \nl
 m_3^2 &=& |d|^2 \l(x^2+y^2+1-2x y \cos\varphi_{ab}
   -2 x \cos\varphi_a + 2 y \cos\varphi_b\r), \nl
 t_{12} &\equiv& \tan {\theta_{12}} =
 \sqrt{ \Delta + \Sigma \over \Delta - \Sigma},
\label{eq:m2}
\eea
where $x \equiv |a|/|d|$, $y \equiv |b|/|d|$,
$\varphi_{ab} \equiv \varphi_a - \varphi_b$ and
\bea
  \Lambda &=& 2 x^2+ 5 y^2+5 + 2x y \cos\varphi_{ab}
   + 2 x \cos\varphi_a - 2 y \cos\varphi_b, \nl
  \Sigma &=& -y^2 + 3 - 2 x y \cos\varphi_{ab} + 6 x \cos\varphi_a + 2 y \cos \varphi_b, \nl
  \Delta &=& \sqrt{\Sigma^2 + 8 (\epsilon_1^2 + \epsilon_2^2)},
\label{eq:Lam_Del}
\eea
where $\epsilon_1 = y^2 + 2x y \cos\varphi_{ab} + y \cos\varphi_b$
and $\epsilon_2 = 3 y \sin \varphi_b$.

We can show that $\Lambda \ge \Delta$ (the equality sign holds for 
$x = 27/32, y=(91-\sqrt{4641})/128$, and $\varphi_a=\varphi_b=0$ ).
So the smallness of the solar mass difference, 
$\Delta m^2_{\odot} = \Delta m_{21}^2 = |d|^2 \Delta$, implies
$\Delta \approx 0$. And this leads to
$\Sigma \approx \epsilon_1 \approx \epsilon_2 \approx 0$.
From this we get a ``magic relation'' for $y = {\cal O}(1)$:
\bea
 && \varphi_b \approx 0, \quad y+ 2 x \cos\varphi_a +1 \approx 0, \nl
 &{\rm or}& \varphi_b \approx \pi, \quad y- 2 x \cos\varphi_a -1 \approx 0.
 \label{eq:magic}
\eea
We note that the above conditions give $\Delta \approx -3 \Sigma$ or 
$\tan\th_{12} =1/\sqrt{2}$.
Then we obtain the tribimaximal mixing (\ref{eq:tribimaximal}) without
further conditions\footnote{If (\ref{eq:magic}) holds exactly, 
$t_{12}$ becomes undetermined.} .
Although the magic relation (\ref{eq:magic}) is not guaranteed by any symmetry, 
it is satisfied
in a large parameter space, especially when the CP violating phase 
$\varphi_a$ is allowed.
In this approximation we get the following mass-squared differences:
\bea
  \Delta m_{21}^2 &\equiv& m_2^2 - m_1^2 \approx |d|^2 \Delta, \nl
  \Delta m_{32}^2 &\equiv& m_3^2 - m_2^2 \approx  6 |d|^2 y \; ({\rm for} \; \varphi_b=0), \;\;
                           {\rm or}\;\;   -6 |d|^2 y \; ({\rm for} \; \varphi_b=\pi).
  \label{eq:msq}
\eea
Here $\Delta m_{21}^2>0$ by definition and it can be identified with the $\Delta m_{\rm sol}^2$.
However, the $\Delta m_{32}^2>0$ can be either
positive or negative:
$\Delta m_{32}^2 = \Delta m_{\rm atm}^2 >0$ for $\varphi_b \approx 0$ (normal
hierarchy) and
$\Delta m_{31}^2 = -\Delta m_{\rm atm}^2<0$ for $\varphi_b \approx \pi$ (inverted
hierarchy).

\begin{figure}[tbh]
\begin{center}
 \includegraphics[width=0.7\textwidth]{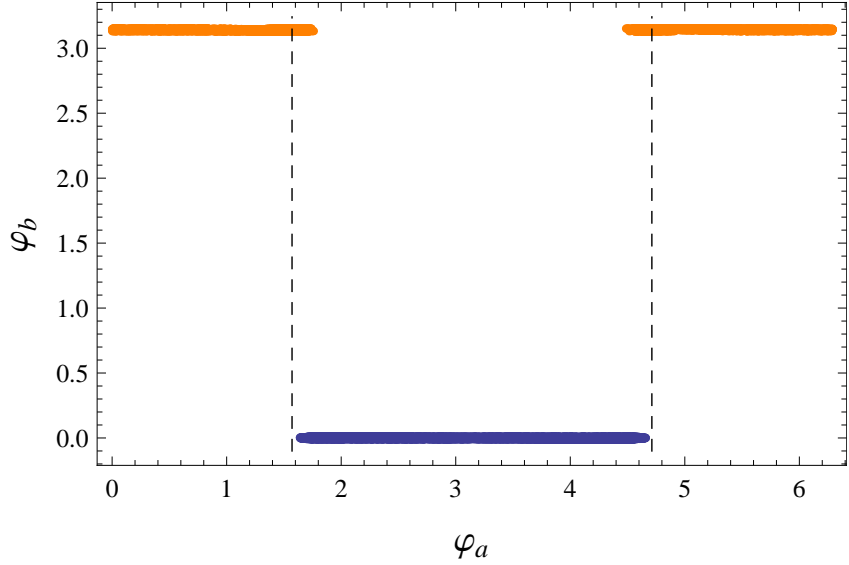}
\end{center}
\caption{Scatted plot in the $(\varphi_a,\varphi_b)$ plane. The blue (orange) color represents 
the case for the normal (inverted) hierarchy. The dashed black lines are $\varphi_a = \pi/2, 3 \pi/2$. }
\label{fig:phiA-phiB}
\end{figure}

Now we get the numerical constraints on the five parameters,
$|d|, x, y, \varphi_a$ and $\varphi_b$,
from the experimental data in Table~\ref{tab:osc_data}.
In Fig.~\ref{fig:phiA-phiB}, we show a scattered plot in the
$(\varphi_a,\varphi_b)$ plane. The blue (orange) color represents the case for the normal (inverted) hierarchy.
We can see the allowed range of $\varphi_b$ 
is quite restricted and looks almost like line (its 
thickness is about 0.02).
This shows the magic relation (\ref{eq:magic}) works quite well.
For the normal hierarchy ($\varphi_b \approx 0$) $\varphi_a$ is restricted in the region $(\pi/2, 3 \pi/2)$, since the $\cos \varphi_a \approx -(y+1)/2x <0$.
 by the magic relation.
For the inverted hierarchy ($\varphi_b \approx \pi$), 
since $\cos \varphi_a \approx (y-1)/2x$, both signs are allowed
in principle. However, the region in which both $x$ and $y$ are small
is excluded 
and the minimum value of $\cos\varphi_a$ allowed
by the data is about $-0.22$.

\begin{figure}[tbh]
\begin{center}
 \includegraphics[width=0.4\textwidth]{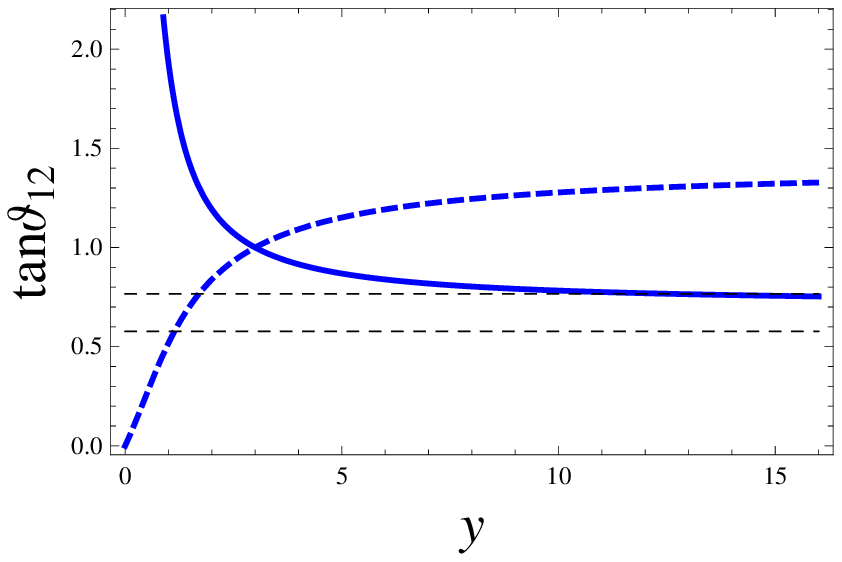}
 \includegraphics[width=0.4\textwidth]{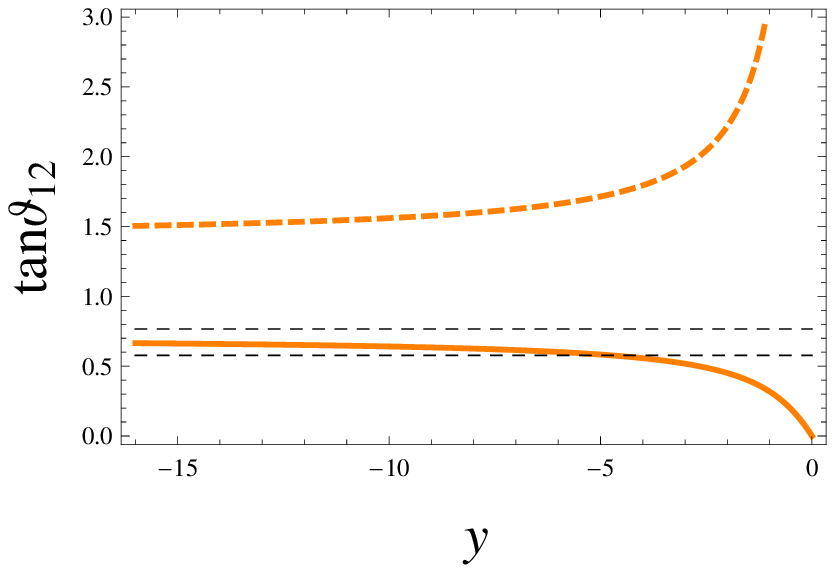}
\end{center}
\caption{ Thick solid (dashed) curve: $\tan\theta_{12}$ as a function 
of $y$ for $y+2 x \cos\varphi_a +1 >0 \; (<0)$.
Dashed horizontal lines: 3-$\sigma$ allowed range of 
$\tan\theta_{12}$, $0.577 < \tan\theta_{12} < 0.766$.
The left (right) panel is for normal (inverted) hierarchy.}
\label{fig:y-t12}
\end{figure}

The very restricted range of $\phi_b$ signifies a fine-tuning. To avoid the fine-tuning problem,
We set $\phi_b =0, \pi$ identically for the normal and inverted 
hierarchy, respectively. To simplify expressions,
we allow $y$ to take negative values from now on, understanding that $y>0$ ($y<0$) implies normal
(inverted) hierarchy.
Then the expressions for the mixing angle $\th_{12}$ simplify greatly and give
\bea
\tan 2 \th_{12} = \pm \frac{2 \sqrt{2} |y| }{y-3} 
\;\;\; {\rm for} \;\; {\rm sign}(y+2 x \cos\varphi_a +1) = \pm.
\label{eq:tan2th12}
\eea
It is interesting to note that this expression is a function of $y$ only and independent of $x$ and $\varphi_a$.
The 3-$\sigma$ allowed range of $y$ can be read from Fig.~\ref{fig:y-t12} and is given by
\bea
  &&  y > 12.7 \;\; {\rm for}\;\; (y+2 x \cos\varphi_a +1) >0 \nl
  && 1.14 <  y < 1.70 \;\; {\rm for}\;\;(y+2 x \cos\varphi_a +1) <0
\eea
for normal hierarchy and
\bea
 y< -4.71 \;\;  {\rm for}\;\; (y+2 x \cos\varphi_a +1) <0
\eea
for inverted hierarchy. 
It is impossible to satisfy the data for the solar mixing angle when  
$(y+2 x \cos\varphi_a +1) >0$.

\begin{figure}[tbh]
\begin{center}
 \includegraphics[width=0.7\textwidth]{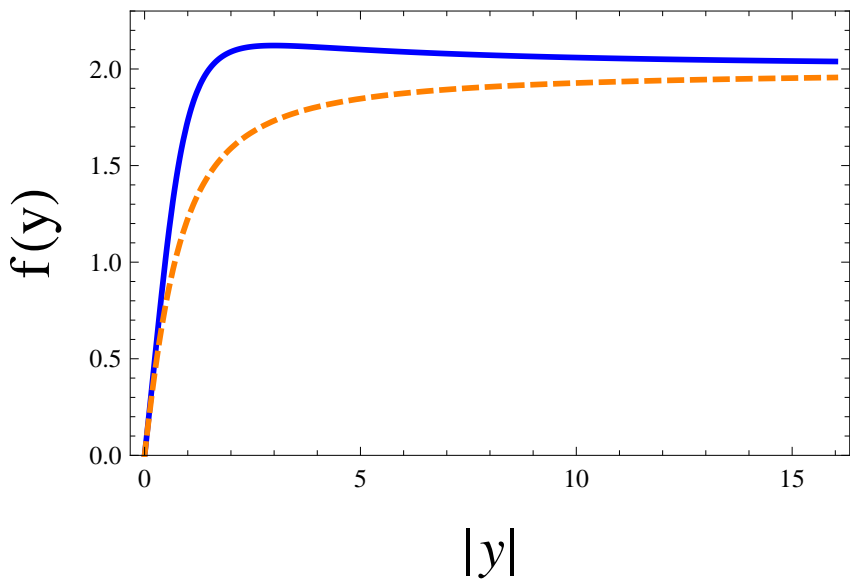}
\end{center}
\caption{Solid (Dashed) curve: $f(y)$ for normal (inverted) hierarchy.}
\label{fig:y-f}
\end{figure}

To consider the constraints from the mass-squared differences, 
we take the ratio 
$|\Delta m_{32}^2|/\Delta m_{21}^2$ because $|d|^2$ is canceled in 
this case.
Then Eqs. (\ref{eq:m2}) and (\ref{eq:Lam_Del}) give
\bea
 \rho \equiv \frac{\Delta m_{\rm sol}^2}{\Delta m_{\rm atm}^2} =\frac{\Delta m_{21}^2}{|\Delta m_{32}^2|}
=\frac{|y+2x \cos\varphi_a+1|}{f(y)}, 
\eea
where
\bea
f(y)= \frac{6 |y|}{\sqrt{9 - 6 y + 9 y^2}}.
\eea
Fig.~\ref{fig:y-f} shows the plots of $f(y)$.

The experimental value of $\rho = (3.19 \pm 0.19) \times 10^{-2}$ 
can be satisfied by tuning $y+2x \cos\varphi_a+1 \approx 0$.

\begin{figure}[tbh]
\begin{center}
 \includegraphics[width=0.7\textwidth]{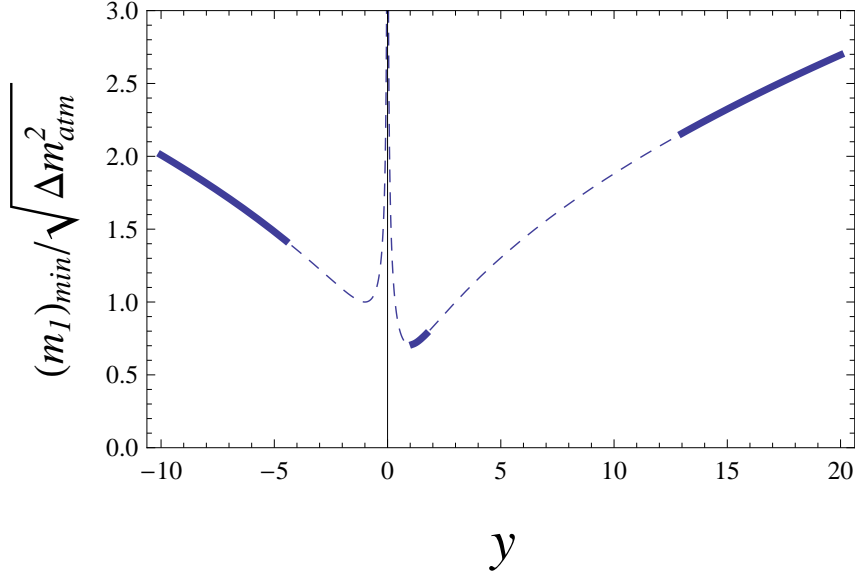}
\end{center}
\caption{The mass ratio, ${(m_1)_{\rm min}}/{\sqrt{\De m_{\rm atm}^2}}$, as a function of $y$ 
when $\cos^2\varphi_a=1$. The thick solid curves are allowed by the current experimental data.}
\label{fig:m1_y}
\end{figure}

Using the relation, $2 x \cos \varphi_a + y + 1 \approx 0$, 
we can get 
\bea
 \frac{m_1^2}{\De m_{\rm atm}^2} \simeq
  \frac{1}{6 |y|}\l(\l(\frac{y+1}{2 \cos\varphi_a}\r)^2+2 (y^2 - y + 1)\r),
\label{eq:m1_atm}
\eea
in the leading order in $\rho$.
The minimum of $m_1$ is obtained when $\cos^2\varphi_a =1$ for both
normal and inverted hierarchy. Fig.~\ref{fig:m1_y} shows 
 ${(m_1)_{\rm min}}/{\sqrt{\De m_{\rm atm}^2}}$, that is, the 
 ${m_1}/{\sqrt{\De m_{\rm atm}^2}}$ as a function of $y$ when $\cos^2\varphi_a =1$.
It shows that there is  a lower bound on the lightest neutrino mass,
$m_{1(3)} \gtrsim 0.03 (0.05)$ eV for the normal (inverted) hierarchy.

\begin{figure}[tbh]
\begin{center}
 \includegraphics[width=0.7\textwidth]{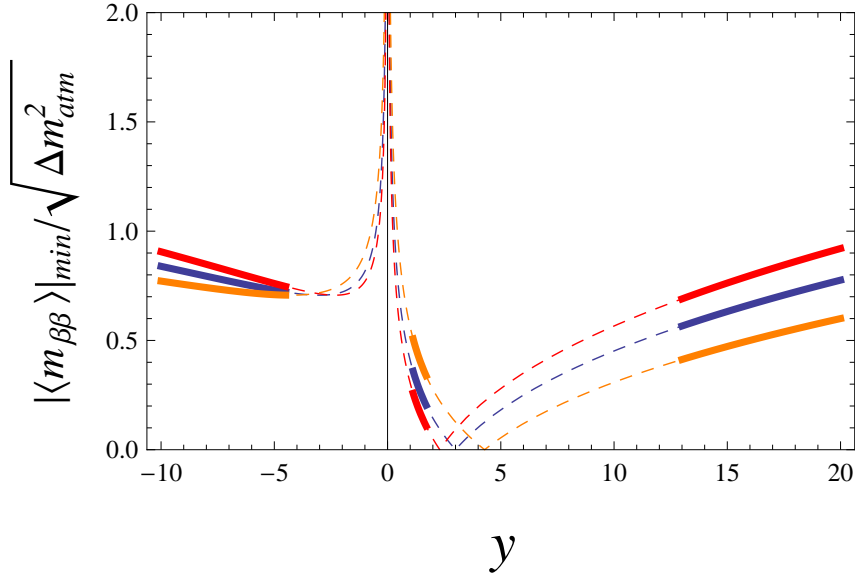}
\end{center}
\caption{The mass ratio, ${(m_{\be\be})_{\rm min}}/{\sqrt{\De m_{\rm atm}^2}}$, as a function of $y$ 
when $\cos^2\varphi_a=1$. The blue (red, orange) curve corresponds to the case
$b=c$ ($\xi=0.3$, $\xi=-0.3$). The thick parts are allowed by the current experimental data.}
\label{fig:mbb-y-xi}
\end{figure}

 Now let us consider the bound on the effective Majorana mass for neutrinoless double-beta
 decay \cite{Hirsch}. 
The amplitude of the neutrinoless double beta decay ($0\nu\beta\beta$) is proportional to
the effective Majorana mass for the $0\nu\be\be$ defined by
 \bea
    \l|\langle m_{\be\be} \rangle \r|
\equiv \l| \sum_i m_i \l(U_{\rm MNS}^*\r)^2_{1i} \r|
 =  \l| \l(M_\nu^{\rm eff}\r)_{11} \r| = \l| a + 2 d\r|
 = d \l| x e^{i \varphi_a} + 2 \r|.
 \eea
Similar to (\ref{eq:m1_atm}), we get 
\bea
 \frac{\l|\langle m_{\be\be} \rangle \r|^2}{\De m_{\rm atm}^2} \simeq
  \frac{1}{6 |y|}\l(\l(\frac{y+1}{2 \cos\varphi_a}\r)^2-2 (y - 1)\r),
\label{eq:mbb_atm}
\eea
in the leading approximation in $\rho$.
The ratio  ${\l|\langle m_{\be\be} \rangle \r|}/{\sqrt{\De m_{\rm atm}^2}}$ has lower bounds
for the allowed range in $y$ (thick parts in the solid blue curve in Fig.~\ref{fig:mbb-y-xi}).
Numerically we get
\bea
&& {\l|\langle m_{\be\be} \rangle \r|}/{\sqrt{\De m_{\rm atm}^2}} > 0.2\; ({\rm for}\; 1.13<y<1.7), \quad
  > 0.57\; ({\rm for}\; y>13),  \nl
&&
 {\l|\langle m_{\be\be} \rangle \r|}/{\sqrt{\De m_{\rm atm}^2}} > 0.72\; ({\rm for}\; y<-4.5).
\eea
The model with $A_4$ symmetry which has non-vanishing lower bound for 
${\l|\langle m_{\be\be} \rangle \r|}/{\sqrt{\De m_{\rm atm}^2}}$
even in the case of normal hierarchy was considered in \cite{Hirsch}.
Numerically we get similar values to theirs.

\subsection{The case $b \not=c$}
\label{sec:bnec}
As mentioned in Section~\ref{sec:2}, the case $b=c$ is not guaranteed by any symmetry.
In this section we extend to the case $b \not=c$.
The effective neutrino mass matrix given in (\ref{eq:nu_eff}) is now in the form:
\bea
  M_\nu^{\rm eff} &=& U(\om)^* M_\nu U(\om)^\dagger =\left(
 \ba{ccc}
  a+2 d &  c   &   b \\
   c    &  b   & a -d \\
   b    &  a-d & c
 \ea
 \right).
\label{eq:nu_eff_full} 
\eea
Since $b=c$ case can already give the tribimaximal mixing matrix, we can see
$b$ and $c$ cannot be so different. Therefore, we can apply the time-independent perturbation
theory to diagonalize the neutrino mass matrix and expand in powers of 
$\xi \equiv (c-b)/2b$. Since $b$ is assumed to be real, we also assume $\xi$ to be real.

Since the 1st and 2nd eigenvalues giving the $\De m_{\rm sol}^2$ are quasi-degenerate, 
the blind application
of the perturbation formula gives unreasonable results. To evade this problem 
we diagonalized the $2 \times 2$ sub-matrix exactly.
And then we applied the perturbation formula in the basis where the first two mass-squared
eigenvalues are diagonal.

The solar mass-squared difference is obtained to be
\bea
\De m_{\rm sol}^2 \simeq d^2 |2 x \cos\varphi_a + y + 1 +\xi y| \sqrt{9 y^2 -6 y + 9}.
\eea
This implies the ``magic relation'' corresponding to (\ref{eq:magic}) is simply replaced by
\bea
 2 x \cos\varphi_a + y + 1 +\xi y \approx 0.
\eea
The atmospheric mass-squared difference is
\bea
\De m_{\rm atm}^2 \simeq  6 d^2 |y| (1 + \xi).
\eea
Since $\xi \ll 1$,  $y>0 \; (y<0)$ still gives normal (inverted) hierarchy.

\begin{figure}[tbh]
\begin{center}
 \includegraphics[width=0.4\textwidth]{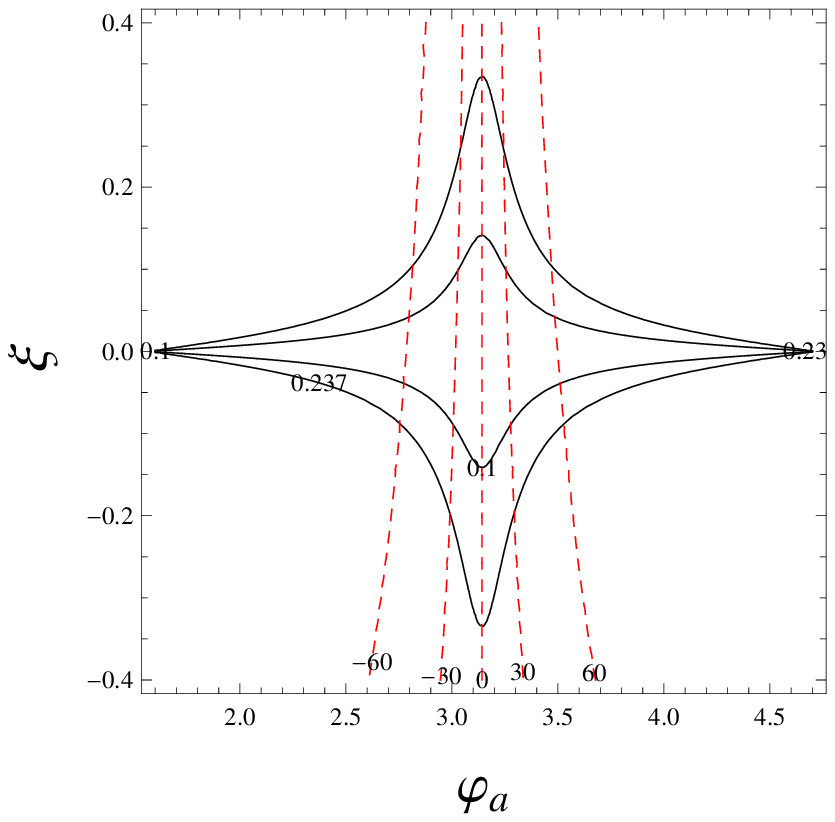}
 \includegraphics[width=0.4\textwidth]{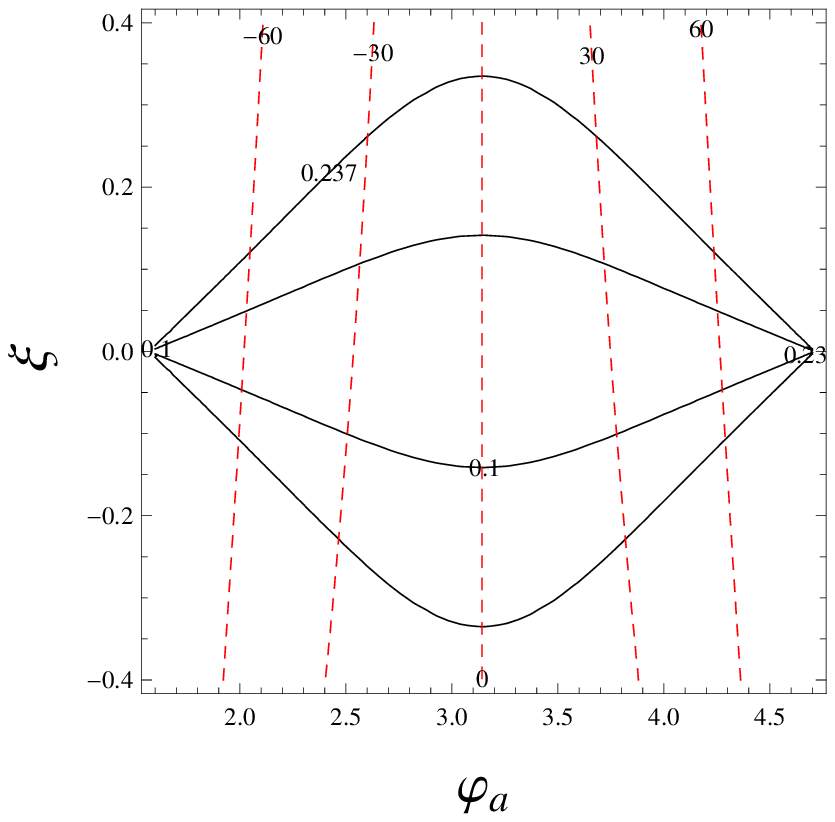}
\end{center}
\caption{Solid (dashed) curves: contours for the constant $s_{13}$ ($\delta$ (in degrees) ) for 
the normal hierarchy. The inner (outer) solid lines are contours for the 3-(1-)$\sigma$ values
of   $s_{13}$. We take
$y=13$ ($y=1.5$) for the left (right) panel.
}
\label{fig:s13_N}
\end{figure}

\begin{figure}[tbh]
\begin{center}
 \includegraphics[width=0.45\textwidth]{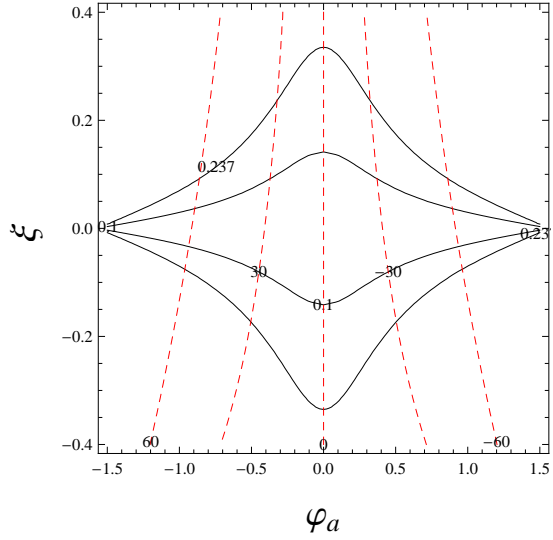}
\end{center}
\caption{The same with Fig.~\ref{fig:s13_N}.
But we take $y=-5$.
}
\label{fig:s13_I}
\end{figure}

The correction in $\xi$ to the effective mass for the $0\nu\be\be$, (\ref{eq:mbb_atm}), is given
by
\bea
 \frac{\l|\langle m_{\be\be} \rangle \r|^2}{\De m_{\rm atm}^2} \simeq
  \frac{1}{6 |y| (1+\xi)}\l(\l(\frac{y+1}{2 \cos\varphi_a}\r)^2-2 (y - 1)
+2 y \xi {y+1 \over (2 \cos\varphi_a)^2}\r).
\label{eq:mbb_atm_xi}
\eea
The minimum values in $\varphi_a$ are obtained for $\varphi_a = \pi (0) $ for normal (inverted)
hierarchy. These are plotted in Fig.~\ref{fig:mbb-y-xi} as a function of $y$ for different
values of $\xi =0, \pm 0.3$. The allowed regions are drawn in thick lines. 
The value $|\xi| \simeq 0.3$ is almost maximum allowed by the 3-$\sigma$ range in $s_{13}$ 
(see Figs.~\ref{fig:s13_N},\ref{fig:s13_I}).
Numerically the minimum values for the normal hierarchy are given by
\bea
  \l({\l|\langle m_{\be\be} \rangle \r|}/{\sqrt{\De m_{\rm atm}^2}} \r)_{\rm min} = 0.1, 0.2, 0.34 \; 
({\rm for}\; \xi=0.3, 0, -0.3, \; {\rm resp.}).
\eea
Since $\rho \ll 0.1$, the correction to (\ref{eq:mbb_atm_xi}) in $\rho$ does not change the results much.

For the mixing angle $\th_{12}$ we get a formula similar to (\ref{eq:tan2th12}),
\bea
\tan 2 \th_{12} \simeq \pm \frac{2 \sqrt{2} |y| }{y-3} 
\;\;\; {\rm for} \;\; {\rm sign}(y+2 x \cos\varphi_a +1+\xi y) = \pm.
\label{eq:tan2th12_full}
\eea

The most significant change from the case $b=c$ is that 
the non-vanishing $\th_{13}$ and consequently $\delta$ for non-trivial
$\varphi_a$ is allowed in $b\not= c$ case. The expression for $s_{13} \equiv \sin\th_{13} $
is obtained by 
\bea
 s_{13} \simeq  \frac{|\xi|}{\sqrt{2}} 
   \sqrt{1 + \l( (y+1+\xi y) \tan\varphi_a \over 3\r)^2}.
\label{eq:s13}
\eea
The CP violating phase $\de$ is given by
\bea
\de \simeq \tan^{-1} \l((y+1+\xi y) \tan\varphi_a \over 3\r) \; ({\rm mod} \; \pi).
\label{eq:delta}
\eea
We do not have a definite prediction for $s_{13}$. But the 3-$\sigma$ range in Table~\ref{tab:osc_data}
can be accommodated. Fig.~\ref{fig:s13_N} (\ref{fig:s13_I}) shows contours for the constant $s_{13}$ and
$\delta$ for the normal (inverted) hierarchy case.  We can see the 3-$\sigma$ range for the $s_{13}$ can be
accommodated in the perturbative region for $\xi$. All the possible values of $\delta$ are allowed
by the current experimental values, although relatively small values of $\delta$ are preferred.

A very interesting prediction is for $\th_{23}$. Up to the first order in $\xi$,
the prediction for $\th_{23}$ is still $\pi/4$. The first correction appears in second order in
$\xi$:
\bea
  \sin^2 \th_{23} = 0.5 + {\cal O}(\xi^2) \approx 0.5 \pm 0.01,
\eea
for the 1-$\sigma$ allowed $\xi$  ($\xi \lesssim 0.1$) by $s_{13}$.
Therefore if the experiments would confirm significant deviation from $\pi/4$ for $\th_{23}$,
our model would be ruled out.

\section{Conclusions}
We studied a triplet Higgs model to generate Majorana neutrino masses
and the mixing matrix in the framework of $A_4$ symmetry.
With the assignments of $A_4$ representations given in Table~\ref{tab:qn},
we see that
\begin{itemize}
\item The tribimaximal form of the neutrino mixing matrix can be naturally obtained for $b=c$.
\item There is a lower bound on the lightest neutrino mass: $m_1 \gtrsim 0.03 (0.05)$ eV
for the normal (inverted) hierarchy.
\item There is a lower bound on the effective mass for the neutrinoless double beta decay:
${\l|\langle m_{\be\be} \rangle \r|}/{\sqrt{\De m_{\rm atm}^2}} > 0.2 (0.57)$
for the normal (inverted) hierarchy.
\item For $b \not=c$ case, we can accommodate the data for the $\th_{13}$.
\item Even for $b \not=c$ case, the prediction for the atmospheric mixing angle does not change 
much from $\th_{23} = \pi/4$ and gives $\sin^2 \th_{23} \approx 0.5 \pm 0.01$, 
which can be tested in near future.
\end{itemize}

\vspace*{12pt}
\noindent
{\bf Acknowledgments}
This work was supported in part by the Korea Research Foundation Grant
funded partly by the Korean Government (MOEHRD) No. KRF-2007-359-C00009 and partly by 
Basic Science Research Program through the National Research Foundation of Korea (NRF)
funded by the Ministry of Education, Science and Technology No. 20090090848 (SB) and 
supported partly by the Korea Research Foundation Grant funded by the Korean
Government (Brain Korea 21) No.200803266004(M.C.Oh).

\end{document}